\documentclass[sigconf]{acmart}

\setcopyright{none}

\usepackage{subfigure}
\usepackage{enumitem}
\usepackage{booktabs}
\usepackage{balance}

\newcommand{\mb}{\mathbf}
\newcommand{\E}{\mathcal{E}}
\newcommand{\R}{\mathbb{R}}
\newcommand{\ie}{\textit{i.e.}}
\newcommand{\eg}{\textit{e.g.}}
\newcommand{\vs}{\textit{vs.}}
\newcommand{\cf}{\textit{c.f.}}

\newcommand{\xhdr}[1]{{\noindent\bfseries #1}.}

\copyrightyear{2018}
\acmYear{2018}
\setcopyright{iw3c2w3}
\acmConference[WWW 2018]{The 2018 Web Conference}{April 23--27, 2018}{Lyon, France}
\acmBooktitle{WWW 2018: The 2018 Web Conference, April 23--27, 2018, Lyon, France}
\acmPrice{}
\acmDOI{10.1145/3178876.3186141}
\acmISBN{978-1-4503-5639-8/18/04}
\begin{document}
\title{Community Interaction and Conflict on the Web}

\author{Srijan Kumar}
\affiliation{
\institution{Stanford University, USA}
}
\email{srijan@cs.stanford.edu}
\author{William L. Hamilton}
\affiliation{
\institution{Stanford University, USA}
}
\email{wleif@stanford.edu}
\author{Jure Leskovec}
\affiliation{
\institution{Stanford University, USA}
}
\email{jure@cs.stanford.edu}
\author{Dan Jurafsky}
\affiliation{
\institution{Stanford University, USA}
}
\email{jurafsky@stanford.edu}

\begin{abstract}
Users organize themselves into communities on web platforms. 
These communities can interact with one another, often leading to conflicts and toxic interactions. 
However, little is known about the mechanisms of interactions between communities and how they impact users. 

Here we study intercommunity interactions across 36,000 communities on Reddit, examining cases where users of one community are mobilized by negative sentiment to comment in another community. 
We show that such conflicts tend to be initiated by a handful of communities---less than 1\% of communities start 74\% of conflicts. 
While conflicts tend to be initiated by highly active community members, they are carried out by significantly less active members. 
We find that conflicts are marked by formation of echo chambers, where users primarily talk to other users from their own community.
In the long-term, conflicts have adverse effects and reduce the overall activity of users in the targeted communities.

Our analysis of user interactions also suggests strategies for mitigating the negative impact of conflicts---such as increasing direct engagement between attackers and defenders. 
Further, we accurately predict whether a conflict will occur 
by creating a novel LSTM model that combines graph embeddings, user, community, and text features. 
This model can be used to create early-warning systems for community moderators to prevent conflicts.
Altogether, this work presents a data-driven view of community interactions and conflict, and paves the way towards healthier online communities. 
\end{abstract}

\maketitle

\vspace{-1mm}
\section{Introduction}
\label{sec:intro}

\begin{figure}[t]
\centering
		 \includegraphics[width=\columnwidth]{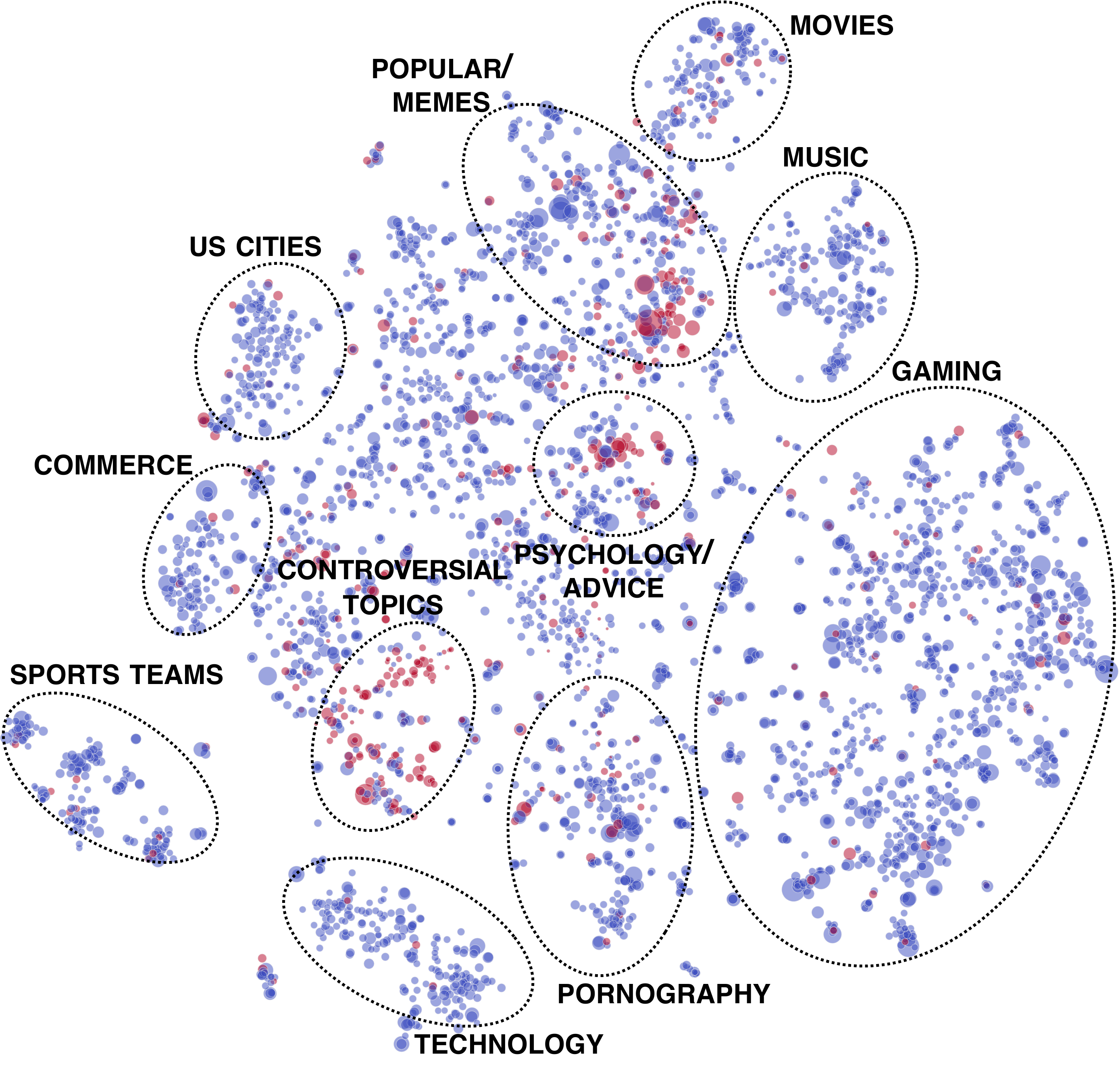}
    \caption{Communities in Reddit: each node represents a community. Red nodes initiate more conflicts, while blue nodes do not. Communities are embedded using user-community information, as described in Section~\ref{sec:predictions}. Figure best viewed in color.
    }
    \label{fig:tsne}
    \vspace{-15pt}
\end{figure}

User-defined communities are an essential component of many web platforms, where users express their ideas, opinions, and share information.
These communities not only provide a gathering place for {\em intracommunity} interactions between members of the same community,
they also facilitate {\em intercommunity} interactions, where members of one community engage with members of another. 
Studies of intercommunity dynamics in the offline setting have shown that intercommunity interactions can be positive---leading to the exchange of information and ideas~\cite{belak2012cross,giles2005intergroup,rahim2010managing,sherif2015group}---or they can take a negative turn, leading to overt conflicts between community members~\cite{hewstone1986contact,university1961intergroup,tajfel_integrative_1979}.
Due to the prevalence of community-level interactions on web platforms, understanding their mechanisms and impact on users is important to foster positive engagement between communities, and to reduce the adverse effects of intercommunity conflicts on users.

However, analyses of intercommunity interaction and conflict are largely absent in previous work on web communities, which tend to focus on community detection~\cite{fortunato2010community,xie2013overlapping}, interactions within individual communities \cite{kraut2012building,leskovec2010predicting,liben2007link}, or how users spread their time between multiple communities \cite{hamilton_icwsm_2017,tan_all_2015,zhang_icwsm_2017}. 
Moreover, existing research in social psychology on conflicts between small groups in lab settings~\cite{hewstone1986contact,sherif2015group,tajfel_integrative_1979} does not give us a window into the fine-grained details of how individual conflicts start on the web, the details of user behavior during these interactions, and their impact on the individuals involved.

A consequence of this methodological limitation is that
we lack the tools to predict and mitigate intercommunity conflicts in large-scale web environments. 
Instances of one online community harassing, ``brigading'', or ``trolling'' another will inevitably occur~\cite{golbeck2017large,hardaker2010trolling,matias2015reporting}, but how can communities defend against these attacks? 
Studies on ``bridging echo chambers''~\cite{garimella2017reducing} and the positive effects of intergroup dialogue \cite{allport1979nature,pettigrew2000does} suggest that direct engagement could be effective for mitigating such conflicts, but there is also ample evidence suggesting that the best defense is simply to ignore the anti-social behavior of the attacking community~\cite{binns2012don}.
Therefore, in web community conflicts, is it more effective to ignore and isolate these attacking users, or is it better to directly engage with them? 
We answer these questions in this work.

\xhdr{Present work}
We provide a large-scale view of intercommunity interactions and conflict through the lens of Reddit, a popular website where users create and participate in interest-based communities called subreddits.
We analyze 40 months of data, containing 1.8 billion comments made by over 100 million users across 36,000 communities \cite{pushshift}.

Naturally, there are no explicit labels of when communities interact with and attack each other.
A key methodological innovation in our work is that we identify and analyze concrete instances of intercommunity interaction and conflict: we identify cases where a post in one ``source'' community hyperlinks to a post in another ``target'' community, and we create a null model of user activity to detect instances where these hyperlinks mobilize a significant number of users to comment in the target post. 
We further use crowd-sourced labels to classify the sentiment of these {\em cross-linking posts} to identify cases of {\em negative mobilization}, where users were explicitly mobilized by posts with negative sentiment. 

To give an example of such negative mobilization, the following post was made on the `r/conspiracy' community, which criticizes and links to a post in `r/Documentaries' community:
\begin{quote}{\small
\textbf{\textit{Come look at all the brainwashed idiots in r/Documentaries}}\\
Seriously, none of those people are willing to even CONSIDER that our own country orchestrated the 9/11 attacks. They are all 100\% certain the ``turrists'' were behind it all, and all of the smart people who argue it are getting downvoted to the depths of hell. Damn shame. Wish people would do their research. The link is [CROSS-LINK]. }
\end{quote}
This post led to several members of r/conspiracy posting angry and uncivil comments on the cross-linked r/Documentaries' post.

\begin{figure}[t]
\centering
		 \includegraphics[width=\columnwidth]{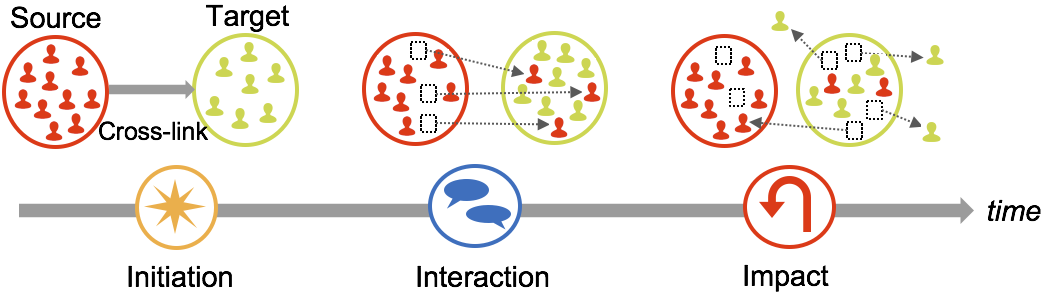}
	\caption{Timeline of interactions: community interactions can be divided into three phases: (i) initiation, where a cross-link is created from the source to the target community, (ii) interaction, where members of the source interact with those in the target, and (iii) long-term impact, where some source members may stay in the target and some target members may leave.}
    \label{fig:timeline}
    \vspace{-12pt}
\end{figure}

Here we provide a data-driven view of these negative mobilizations, which can broadly be divided into three phases (Figure~\ref{fig:timeline}).
The first phase is {\em initiation}: a user in the source community makes a cross-link to the target community that (potentially) mobilizes a subset of users (Figure~\ref{fig:timeline}, left). 
The second phase is {\em interaction}: once the cross-linking post has mobilized a subset of users, these attackers begin to comment on the comment thread of the target post and interact with members of the target community (\ie, the defenders; Figure~\ref{fig:timeline}, middle). 
The final phase is {\em impact}: even after the negative mobilization is over, the event may have long-term impacts on the behavior of the attacking and defending users (Figure~\ref{fig:timeline}, right). 
By analyzing how thousands of negative mobilizations proceed through these three phases, we characterize the types of users and communities who initiate and participate in intercommunity conflicts, and highlight possible mitigation strategies that defending communities could use to minimize their adverse effects.

\xhdr{Initiation of mobilizations}
Our findings show that a small set of communities is responsible for the majority of negative mobilizations: 74\% of negative mobilizations are initiated by 1\% of source communities.
Moreover, we find that these interactions generally occur between highly similar communities. 
Figure~\ref{fig:tsne} shows a learned, two-dimensional social map of the Reddit communities (see Section \ref{sec:predictions} for details), where each node is a community and its redness indicates the proportion of cross-links it generates with negative sentiment. 
The relatively few conflict-initiating (\ie, red) communities are concentrated in three dense regions of this Reddit social map.
Surprisingly, at the user level, we find that while negative mobilizations are initiated by highly active members of the source community, the actual perpetrators of negative mobilizations are users that are significantly less active.

\xhdr{User interactions during mobilizations}
By analyzing the user-user replies on the target threads during negative mobilizations, we gain fine-grained insight into dynamics of user behavior in intercommunity conflicts. 
We create two variants of the PageRank algorithm, which measure the flow of information from the perspective of the attackers and defenders, respectively.
Using these measures, we find that an important marker of negative mobilizations is an echo chamber effect, where participants preferentially interact with members of their ``home'' community, \ie, attackers talk to other attackers, and defenders to other defenders. 
Furthermore, we find that attackers tend to single out and collectively ``gang-up'' on a small set of defenders.

\xhdr{Impact of mobilizations}
Negative mobilizations have a long-term adverse impact on the involved users and communities.
We find that negative mobilizations lead to a ``colonization'' effect, where 
defenders reduce their participation in the target community while 
attackers become more active.
However, building off our analysis of user-level interactions, our results suggests a possible route for mitigating this adverse outcome: we show that a reduction in the echo-chamber effect---marked by an increased interaction between attackers and defenders---and an increase in defenders' use of anger words towards attackers, are associated with decreased rates of colonization and increased future participation rates for the defenders. 
Thus, it appears that increased engagement and a more fierce defense may be a more effective mitigation strategy, compared to ignoring or isolating the attacking users.

\xhdr{Predicting mobilizations}
Lastly, we develop models to predict whether a cross-link will lead to a mobilization. 
Our model combines recent advancements in deep learning on social networks with a novel variant of recurrent neural network LSTM model, which we call ``socially-primed'' LSTM model. Our model achieves an AUC of 0.76 in predicting if a cross-link will lead to a mobilization, significantly outperforming a strong baseline based on expert-crafted features, which achieves an AUC of 0.67.
This model could be used in practice to create early warning systems for community moderators to predict potential negative mobilizations as soon as the cross-linking post is created, and help moderators to curb the adverse effects of intercommunity conflicts. 

Altogether, our results shed light on how intercommunity interaction and conflicts occur on the web, and pave the way towards the development and maintenance of healthier online platforms.

\vspace{-1mm}
\section{Data and definitions}
\label{sec:data}

Users on Reddit form and join interest-based communities called subreddits (\eg, `r/Documentaries' or `r/StarWars'), and within these communities they post and comment on content (\eg, images, videos, links to articles, etc.). 
We use 40 months of Reddit post and comment data, from January 2014 to April 2017~\cite{pushshift}, to analyze the interactions between these subreddit communities. 
All our analysis was conducted with publicly available data~\cite{pushshift}, and relevant codes and data are available on our project website~\cite{appendix}.

As there are no explicit labels of intercommunity interaction on Reddit, we leverage hyperlinks where a post in one community (which we call the {\em source}) links to a post in another community (which we call the {\em target}).
These {\em cross-links} initiate a (possible) interaction between the two communities (\ie, the first step \emph{`initiation'} in the timeline of Figure~\ref{fig:timeline}). 
By following the cross-link, users may be \textit{mobilized} from the source to comment on the linked post in target community, thus leading to intercommunity \emph{`interactions'} (step two in Figure~\ref{fig:timeline}).
After removing overlapping cross-links, we obtain 137,113 cross-links made between 36,000 communities, which is the set we analyze.

User activity on Reddit primarily occurs in discussion threads associated with posts.
In order to identify the explicit effect of cross-links and to control for community-level differences and temporal confounds in our analysis of these threads, we use a matching process to provide a reference point for the posts that are involved in a cross-link (\ie, both the source post that contains the cross-link and the target post that is linked to). 
In particular, for each post $p$ that we analyze, we select a {\em matched post} from the same community that was created closest in time to $p$ and that does not have any outgoing or incoming cross-links. 
The temporal restriction ensures that hourly, daily, or weekly changes in user behavior are controlled for, and selecting matched comparison posts from within the respective communities ensures that we are not confounded by community-level differences. 

For the purpose of our analysis, we also define the {\em members} of the source, $S$, and target, $T$, communities at the time of each cross-link.
In particular, for a cross-link that is made on day $d$, we define members of community $S$ (resp., $T$) as users who have made at least one comment in $S$ (resp., $T$) in the 30 days prior to $d$, but who did not comment in $T$ (resp., $S$) during this time period.\footnote{But note that for analyses of user history, we always ignore comments made within ${\pm}3$  days of the cross-link to remove possible confounds.} 
In other words, members are users who participated in community $S$ (resp., $T$) but not in community $T$ (resp., $S$) before the cross-link was made.\footnote{We focus on exclusive members of the two interacting communities, following research conducted in the social psychology literature~\cite{hewstone1986contact,sherif2015group,tajfel_integrative_1979} where participants are always exclusively members of either communities. 
Other types of users, such as common members, conflate the observations of the two communities, while new users or non-members add little additional information. Their effects can be studied in future research.
}
To provide baseline comparisons for these users, we again use a matching process.
For each member of $S$ who comments on the target thread due to mobilization, we sample a random comparison member of community $S$ who did not do so. 
This {\em matched user} is selected such that it made similar number comments in the past 30 days as the source member. An analogous process is used to match members of community $T$.

\vspace{-2mm}
\subsection{Defining mobilizations}
We start by identifying cases of mobilization. 
We define mobilizations as cases where a cross-link leads to an increase in the number of comments by current source members on the discussion thread of the target post. 
Identifying such mobilizations is challenging because simply counting the number of source community members who comment in the target thread ignores that current source members may comment on the target community at random, instead of as an effect of the cross-link.
Therefore, to identify cases where source members are mobilized \textit{due to the cross-link}, we compare against a null model that measures the expected rate of comments of source members on the target thread.
To control for the initial popularity of the posts, the null model is created using comments made on the matched target posts, restricting to cases where the target and matched posts have a near-equal number of comments before the cross-link is made (in particular, where the difference in their comment counts is less than 5). 
We compare the number of comments made by source members on the two threads, within a 12 hour window before and after the cross-link is created.

\begin{figure}[t]
 \vspace{-10pt}
\centering
		\subfigure{
        \includegraphics[width=0.65\columnwidth]{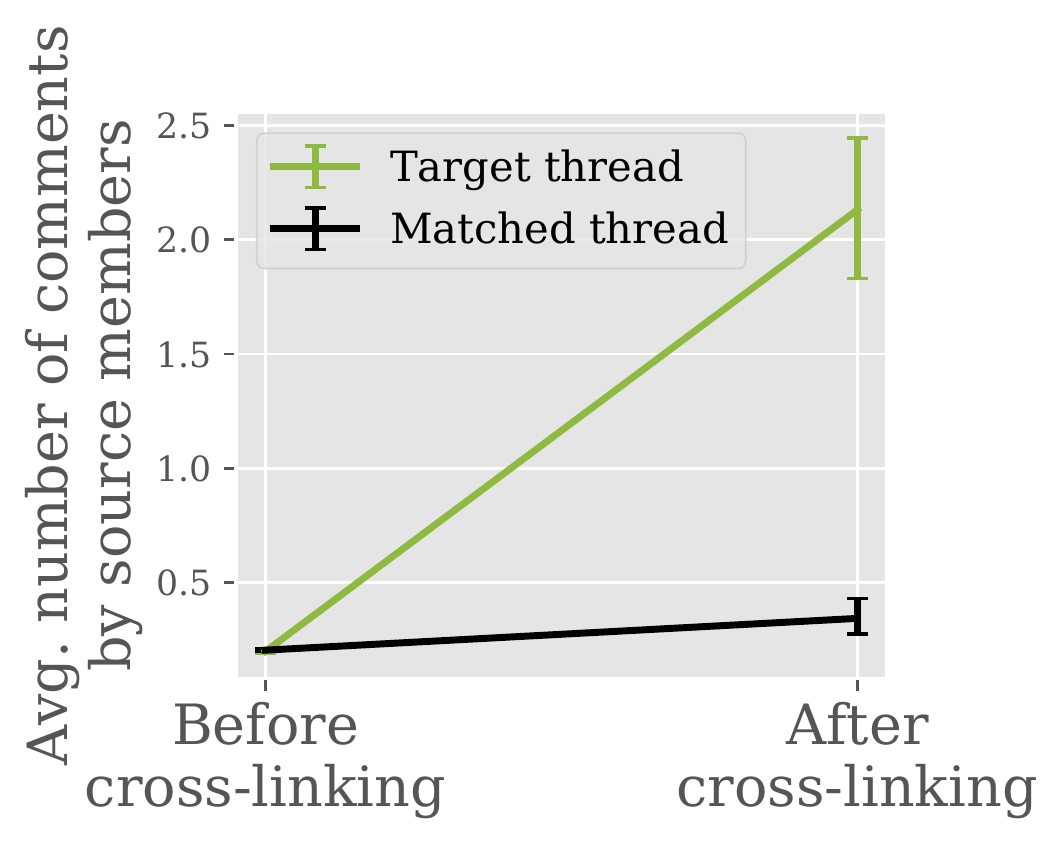}
        }
 \vspace{-10pt}
    \caption{Cross-links lead to an increase in number of comments by source members in the target thread.} 
    \label{fig:definition}
    \vspace{-10pt}
\end{figure}

Figure~\ref{fig:definition} shows the change in average number of comments made by source members in the cross-linked target thread compared to a matched thread, restricting to those pairs whose pre-cross-link comment counts are equal.
However, after cross-linking, there is an increase in both the threads: on average, the matched threads exhibit a $1.6\times$ relative after-to-before increase, while the target threads exhibit an $8.8\times$ increase. 
The baseline $1.6\times$ increase for the matched threads illustrates the expected increase in comments by source members under our null model (\ie, irrespective of cross-links).
Based on these observations, we define {\em mobilizations} as cases where a cross-link leads to an increase in the number of comments by source community members in the target thread that is more than the baseline rate (\ie, a ${>}1.6\times$ after-to-before increase). 
This gives a total of 22,075 mobilizations, which is about 16\% of the total set of cross-link posts we study.

\vspace{-2mm}
\subsection{Classifying the sentiment of mobilizations}
To further categorize interactions, we classify cross-links based on the sentiment of the source post.
Overtly negative intent of the source post towards the post in the target community may signal outgroup derogation~\cite{hewstone2002intergroup}, which is a fundamental component of intergroup conflict~\cite{tajfel_integrative_1979,tajfel1982social}. 

We collected labels from Mechanical Turk crowdworkers, who were shown source and target posts pairs, and were asked to label the sentiment of the source post towards the target as either negative, neutral, or positive.
The workers reported very few instances of positive intent, so we merged the positive and neutral classes. 
We obtained two labels each (negative vs. neutral) for a randomly sampled set of 1020 pairs, with inter-rater agreement of over 0.95.
We then converted the text of the source post\footnote{We removed words common in the source and target posts as source post may quote (part of) the target post.} into a large set of features: lexical features from the LIWC \cite{tausczik_psychological_2010} and VADER \cite{hutto2014vader} lexicons, as well as stylistic linguistic features, such as average word length, readability scores, and punctuation counts.\footnote{The full feature list is available in the online appendix~\cite{appendix}.}
Using a Random Forest classifier, our classification model achieves an accuracy of 0.80 using 10-fold cross validation.\footnote{We used the implementation in scikit-learn package~\cite{pedregosa_scikit_2011} with forests of 400 trees.}
We used this classifier to categorize the remaining source posts, and it labeled 8\% of them as negative.

Coupling this sentiment score with mobilization, we call mobilizations that are initiated with negative sentiment as {\em negative mobilizations} (a total of 1809 cases), and those that start with neutral sentiment as \textit{neutral mobilizations} (a total of 20266 cases). 

Here, we also define two terms: 
\textit{attackers} are the members of the source community, and \textit{defenders} are the members of the target community, who comment in the (cross-linked) target thread during a negative mobilization.
Note that we use these terms as they are clear and intuitively understandable.
However, not all ``attackers'' are necessarily negative while commenting (though we do find that they most often are (\cf\ Section~\ref{sec:interaction})). Similarly, ``defenders'' may not be defending during a negative mobilization as such, but the terminology is used for ease of explanation.

\vspace{-1mm}
\section{Initiation of Mobilizations}\label{sec:initiation}

We start our observations with the first phase, the \textit{initiation} of intercommunity interactions, as shown in Figure~\ref{fig:timeline}.
We aim to characterize the properties of the key entities involved in these interactions: the two communities and their participating members. 

\begin{figure}[t]
\centering	
		\subfigure{	
        \includegraphics[width=0.65\columnwidth]{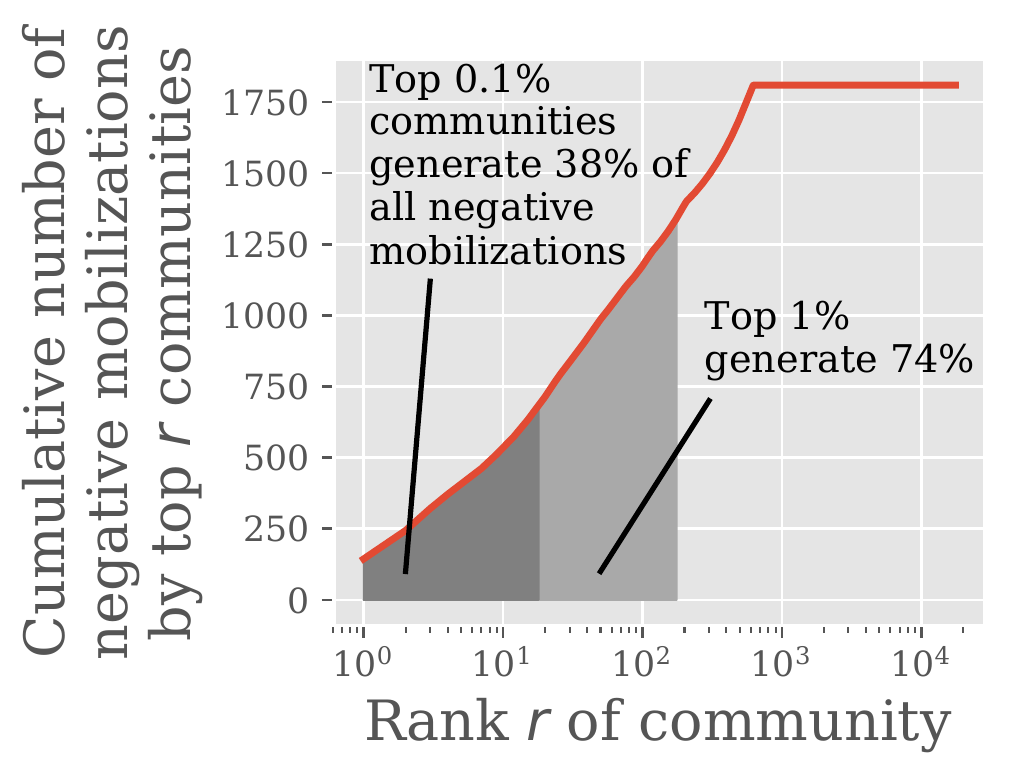}
        }
		 \vspace{-15pt}
    \caption{A small number of communities initiate the majority of conflicts.}
    \vspace{-12pt}
    \label{fig:conflict-sources}
\end{figure}

\vspace{-1mm}
\subsection{Which communities initiate negative mobilizations?}
Contrary to in-lab studies which bring together two groups to elicit interactions and intergroup conflict~\cite{hewstone1986contact,sherif2015group,tajfel_integrative_1979}, web-based platforms have thousands of communities that could potentially interact. So, among these communities, which tend to initiate negative mobilizations, and which communities do they target?

We start by looking at the properties of the initiating community.
We find that most negative mobilizations are initiated by a small number of communities (see Figure~\ref{fig:conflict-sources})---less than 0.1\% and 1\% of source communities are responsible for 38\% and 74\% of the negative mobilizations, respectively.
This means that these handful of communities are hubs of negatively mobilizing users.
This finding echoes those on anti-social behavior, which show that troll-like behaviors are concentrated in small number of communities~\cite{cheng2015antisocial}, and that taking precautionary measures, such as banning particularly egregious communities, can be effective in curbing this behavior~\cite{chandrasekharan2017you}.

Next, we look at the similarity between the source and target communities involved in mobilizations. 
Previous work has suggested that conflicting communities are likely to focus on topically similar areas~\cite{datta2016misaligned,hamilton2016inducing}, so we evaluate this hypothesis here. 
Looking at the pair of interacting communities, we quantify the similarity of two communities as their tf-idf post similarities.\footnote{The tf-idf similarity value~\cite{salton1988term} is computed over all communities posts using the top-10,000 words in terms of overall frequency-rank on Reddit.}
We find that negative mobilizations (and mobilizations in general) tend to occur between highly similar communities---the average tf-idf content similarity between the source and target communities in mobilizations is $1.5\times$ higher than the average value computed between random community pairs (0.51 \vs\ 0.34; $p < 0.001$).\footnote{All reported p-values are based on the Wilcoxon signed-rank test for paired comparisons and the Mann Whitney U-test otherwise ~\cite{siegal1956nonparametric}.}
Manually examining these community pairs reveals that they tend to focus on similar topics, but have different views on the subject matter (\eg, r/conspiracy \vs\ r/worldnews, or r/mensrights \vs\ r/againstmensrights), which fits with previous discussions of this phenomenon \cite{datta2016misaligned,hamilton2016inducing}.

\begin{figure}[t]
\centering
		\subfigure{
        \includegraphics[width=0.40\textwidth]{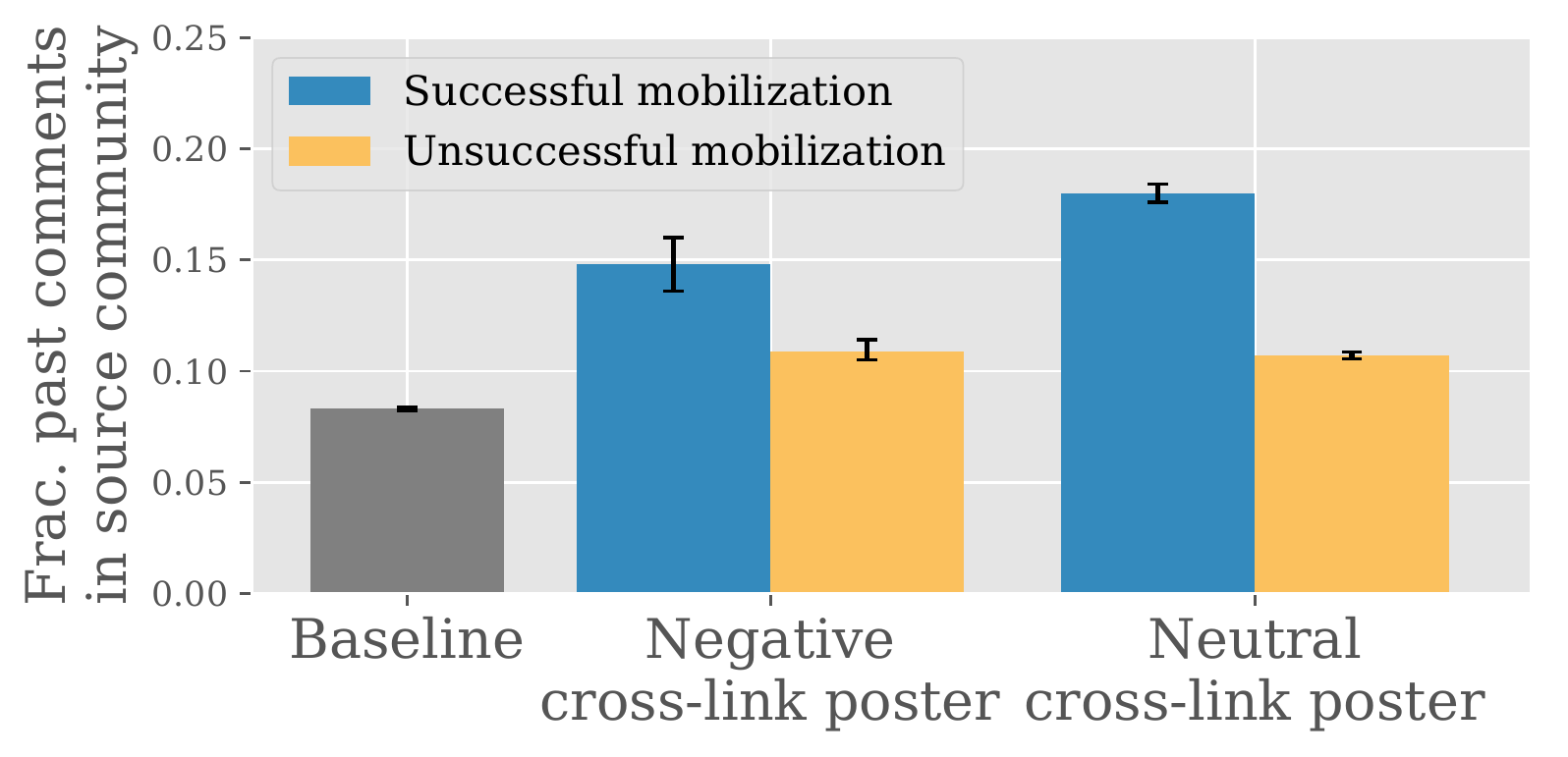}
        }
        \vspace{-10pt}
    \caption{Successfully negative mobilizing posts are created by highly active `core' members of the source community.}
    \label{fig:conflict-users}
     \vspace{-10pt}
\end{figure}

\vspace{-1mm}
\subsection{Which users are involved in negative mobilizations?}
In web-based communities comprised of thousands or even millions of users---with varying levels of participation---only a small fraction of users will actually be involved in any individual negative mobilization.
So, what are the properties of the users that get involved as opposed to the ones that do not?
We answer this question by looking at the user who creates the cross-link post that initiates the negative mobilization, as well as the users that get mobilized and participate in the resulting discussion.

Examining the cross-link creator, we see that negative posts with cross-links are created by users that are 10\% more active in the source community compared to its random matched user ($p<0.001$, Figure~\ref{fig:conflict-users}(a); see Section~\ref{sec:data} for details on matching). As not all cross-links lead to mobilization, we further see that users that are successful in mobilizing others are significantly more active than the ones who are not ( $p<0.001$). Thus, we see that highly active members of the source community are responsible for initiating negative mobilizations. 
Similar observations are made in the case of neutral mobilizations.

But who are the actual perpetrators of negative mobilizations? Are they also highly active members of the source community?
To answer this, we compare the mobilized attacking source members (the attackers) to their matched counterparts. 
We find that the attackers are, in fact, significantly less active in the source community (fraction of past comments in source community is 0.30 vs. 0.17, $p<0.001$). 
We also find that the attackers expressed more `anger' in their past comments;\footnote{This is calculated using comments made in past 30 days.} attackers use anger words (from the LIWC lexicon \cite{tausczik_psychological_2010}) $1.2\times$ more than their matched counterparts (0.31 vs. 0.26, $p<0.001$).

Similarly, examining the members of the target community who are mobilized to defend during these attacks, we find analogous results---these defenders are less active in the target community (fraction of past comments in target community is 0.27 vs. 0.19, $p<0.001$) and use $2.2\times$ more anger words than their matched counterparts (0.32 vs. 0.145, $p<0.001$).
Thus, we see that the defending and attacking users are similar: they tend to be less active in their home communities and have used more anger words in the past.

\vspace{1mm}
Overall, we find that negative mobilizations are initiated by a handful of communities that attack highly similar communities. 
While these interactions are initiated by the highly active users of the source community, the attackers and defenders who actually get mobilized to participate in the negative mobilization are much less active than them---a finding that dovetails with psychological studies showing that peripheral group members are more likely to engage in public displays of outgroup derogation \cite{noel1995peripheral}. 
Prior work has argued that this phenomenon is due to the desire of peripheral members to increase their ingroup status (\textit{ibid.}), and interestingly we see this effect even on Reddit, which is pseudonymous.

\vspace{-1mm}
\section{User Interaction During Mobilizations}
\label{sec:interaction}

Now that we have characterized the entities involved in negative mobilizations---both in terms of the communities and users involved---we turn to the task of characterizing the dynamics of user interactions during these events. 
Do attackers and defenders talk to each other, or do they create separate bubbles where they talk to other similar users, creating an echo chamber-like effect?

We answer these questions by analyzing the user-to-user reply network of discussions on the target post, in the second phase of the mobilization, \ie, the \textit{interaction} phase (see Figure~\ref{fig:timeline}).
This gives us the unique ability to study thousands of highly rich user-user interactions during negative mobilizations, providing a new degree of both scale and granularity to the study of intercommunity conflict~\cite{tajfel_social_2010}.

\begin{figure}[!t]
\centering      
        \includegraphics[width=0.55\columnwidth]{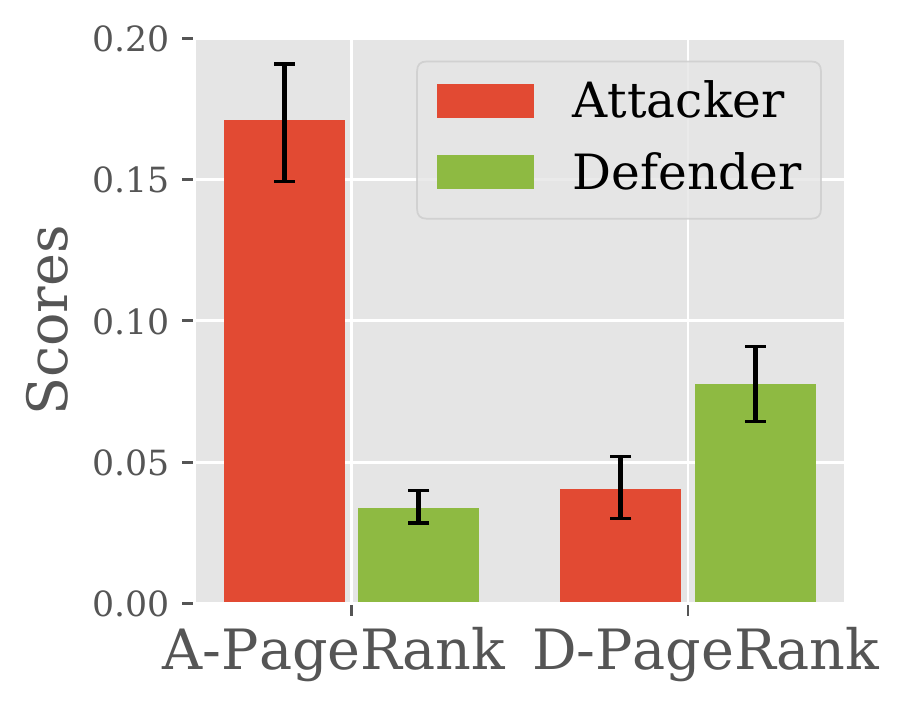}
        \vspace{-7pt}
    \caption{ Echo chambers are formed during negative mobilization: attackers have higher A-PageRank scores than defenders, and defenders have higher D-PageRanks.}
    \label{fig:interactions}
    \vspace{-10pt}
\end{figure}

\xhdr{Negative mobilizations lead to negative interactions}
We find that negative mobilizations have a negative impact on the sentiment and civility of the target discussion thread. 
We compare the comments made on the target threads of negative mobilizations and their matched threads (see Section~\ref{sec:data} for matching details).
The target threads are more likely to contain anger words (44\% increase; $p < 0.001$) and are $25\times$ more likely to have a comment removed by a moderator (deletion rates of 0.205 \vs\ 0.008; $p<0.001$).\footnote{Note that this result is also statistically significant after macro-averaging deletion rates across communities, indicating that it is not simply due to the target communities have higher baseline deletion rates.}
For comparison, these adverse effects are drastically reduced in neutral mobilizations (\ie, no significant change in deletion rate, $p>0.05$).

\xhdr{Quantifying echo chambers in negative mobilizations}\\
Do attackers and defenders talk to each other during negative mobilizations? Or do they occupy separate worlds, thus exhibiting homophily and creating echo chambers?
We answer this question by constructing user-user reply networks from comments made on the target thread.
The network is directed and weighted, where the edge weight from user $i$ to user $j$ corresponds to the number of times $i$ directly replied to one of $j$'s comments.

Our analysis suggests a homophily and echo chamber effect, where attackers preferentially interact with other attackers and defenders with other defenders.
For example, examining the direct interactions between users (\ie, cases where $i$ replied to $j$ or vice-versa), we find that attackers interact $2\times$ more with other attackers compared to their interactions with defenders, while defenders interact $20\times$ more with other defenders compared to attackers ($p<0.001$).
We quantify this echo chamber effect using two variants of the PageRank score~\cite{page1999pagerank}: 
the Defender PageRank (D-PageRank) and Attacker PageRank (A-PageRank), which quantify the centrality of a user in the reply network from the perspective of defenders and attackers, respectively. 
Like the classic PageRank statistic~\cite{page1999pagerank}, the D- and A-PageRank values for a user, $i$, correspond to the probability of a random walk visiting the node corresponding to $i$, where this random walk follows directed edges with transition probabilities proportional to edge weights and ``teleports'' to a new random start point with a probability of 0.25 at each step. 
However, unlike the standard PageRank, we compute the D-PageRank and A-PageRank by restricting the teleport set to defender nodes and attacker nodes respectively, similar to the approach used by Garimella et al. to quantify controversy~\cite{garimella2016quantifying}.

Figure~\ref{fig:interactions} shows the values of these scores. We see the attackers have significantly higher A-PageRanks---meaning that they are more likely to be visited on random walks starting from other attackers---while defenders have significantly higher D-PageRanks.
Thus, we see that the discussion threads during negative mobilizations seem to be marked by a bifurcation between attackers and defenders, where members primarily talk to other members from their ``home'' community.
Similar echo chamber effects have been found during controversial discussions in social media platforms \cite{conover2011political,faris2017partisanship,garcia2012political,ribeiro2017everything}.

\xhdr{Attackers ``gang-up'' on defenders}
Of course---despite the echo chamber effect elucidated above---attackers and defenders will inevitably have some direct interactions. 
But what do these interactions look like?
Examining the defenders' A-PageRank scores, we find that while 83\% defenders have exactly zero score, a very small fraction of defenders (1.14\%) have an A-PageRank score which is at least ten times the mean value among all attackers and defenders.
This highly skewed distribution of the defenders' A-PageRank scores indicates that a small set of defenders are involved in most of the interactions with the attackers.
Additionally, we find that the comments made between attackers and defenders 
employ very angry language---attackers use anger words more frequently when talking to defenders than when talking to fellow attackers (0.015 \vs\ 0.011; $p<0.001$), and vice-versa for defenders (0.017 \vs\ 0.014; $p<0.001$). 
Together, these results suggests that while attackers do not communicate with most of the defenders, they tend to single out and ``gang-up'' on a handful of defenders.

\vspace{2mm}
Overall, we find that negative mobilizations lead to the creation of echo chambers in the target discussion, where attackers primarily talk to other attackers and defenders talk to other defenders.
But whenever they do talk to each other, they exchange highly angry comments.
Moreover, our analysis shows that attackers tend to ``gang-up'' to attack some defending users, while other defending users do not engage with the attackers.

\vspace{-1mm}
\section{Impact of mobilizations}\label{sec:impact}

As seen in the previous section, attackers ``gang-up'' on defending users during negative mobilizations. 
But in the long-term, do these attacks have any impact on the involved users?
And what should the defenders do to prevent the negative effects of these events---should they ignore the attackers during discussions, or should they actively engage with them? 
In this section, we answer these questions by studying the third phase of mobilization (Figure~\ref{fig:timeline}), \ie, the \textit{impact} phase, where we measure changes in user activity after the conflict is over.

\begin{figure}[t]
\centering
         \subfigure{
        \includegraphics[width=0.6\columnwidth]{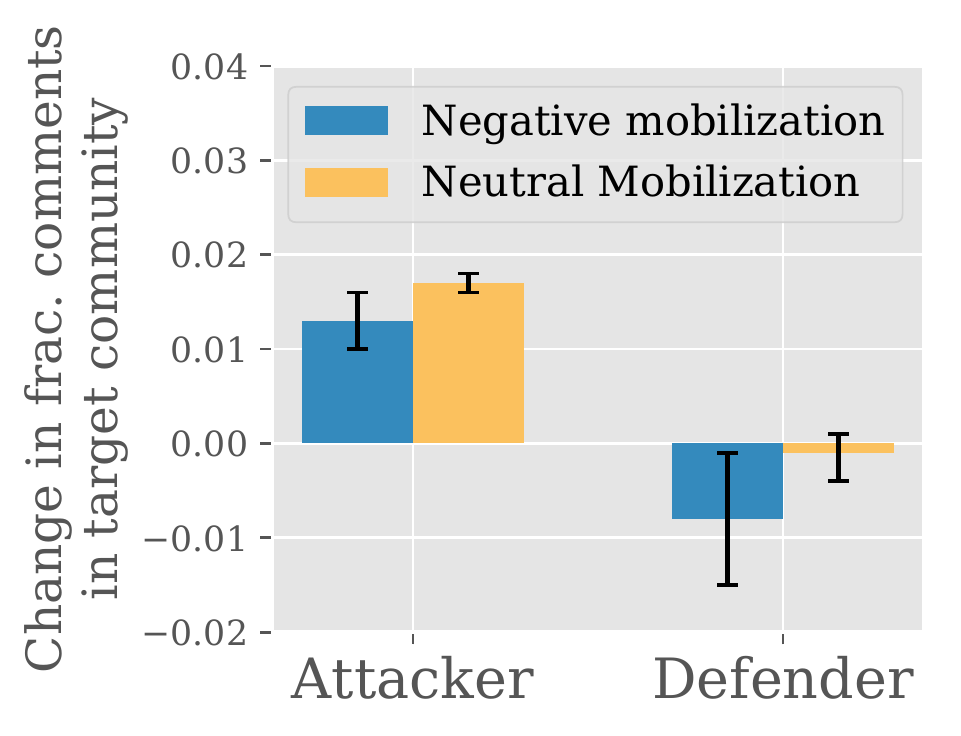}
        }
         \vspace{-10pt}
    \caption{Impact of negative mobilization: defenders become less active in target community, while attackers become more active.}
    \label{fig:impact}
    \vspace{-10pt}
\end{figure}

\begin{figure*}[t]
\centering
		\subfigure[]{
        \includegraphics[width=0.8\textwidth]{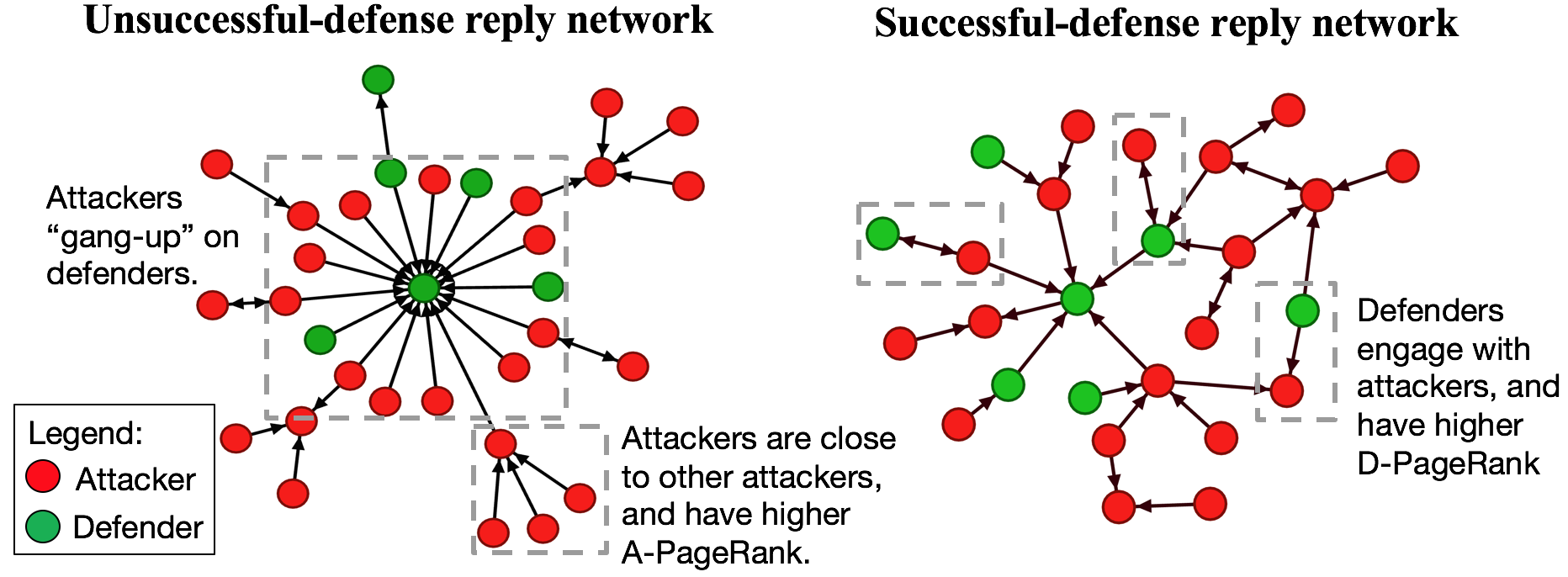}
        }

		\vspace{-2mm}
        \subfigure[]{
        \includegraphics[width=0.24\textwidth]{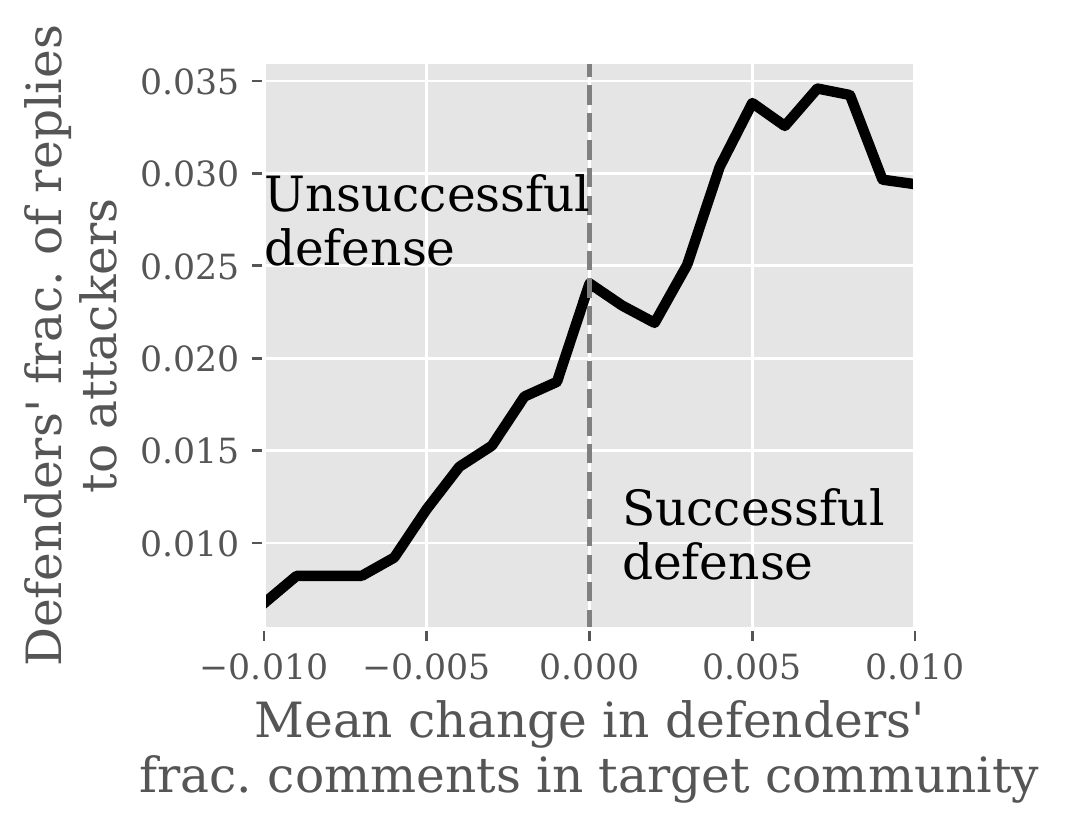}
        }
		\subfigure[]{
        \includegraphics[width=0.23\textwidth]{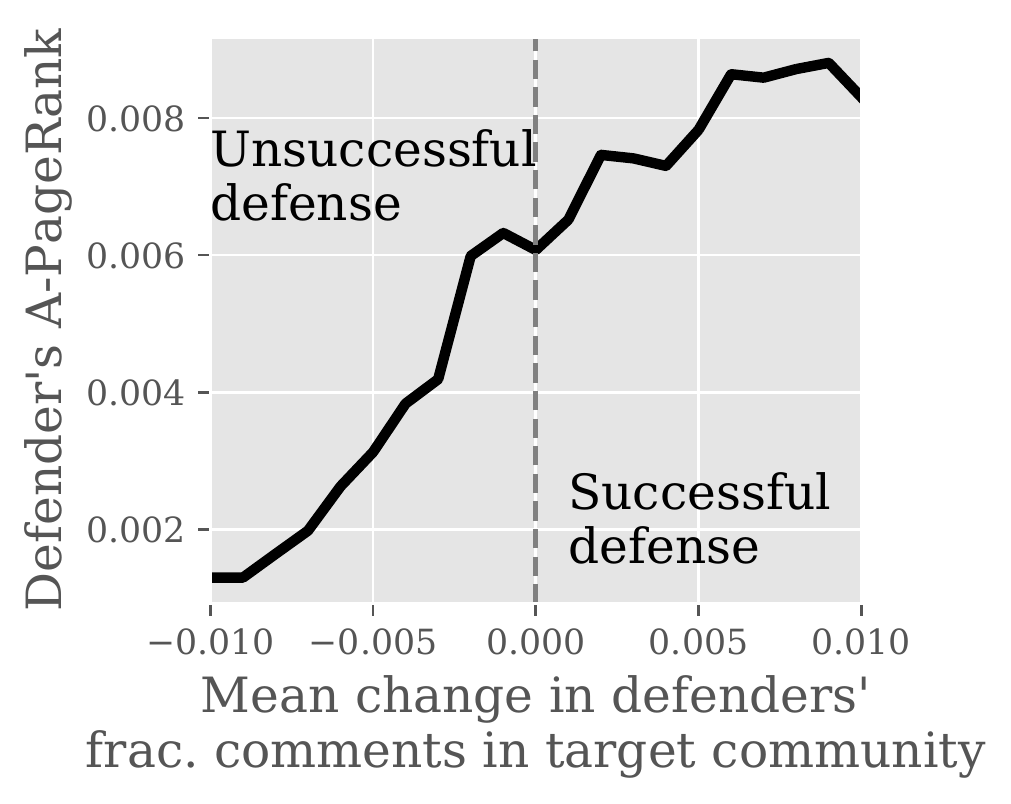}
        }   
        \subfigure[]{
        \includegraphics[width=0.23\textwidth]{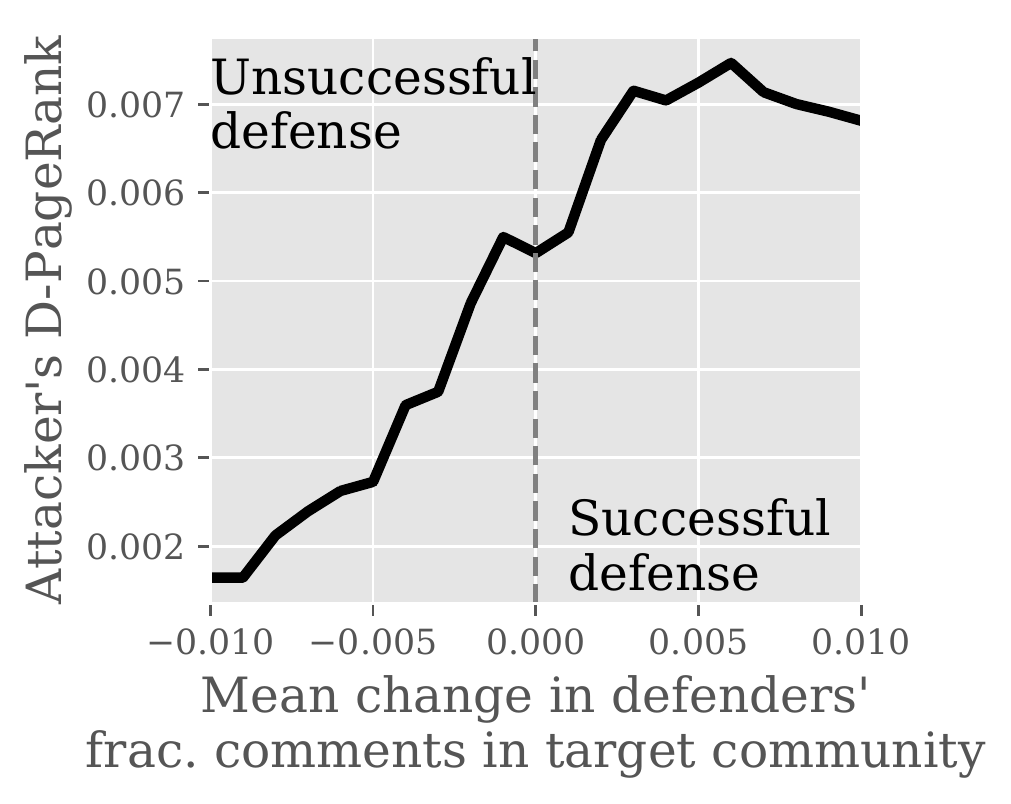}
        }
		\subfigure[]{
        \includegraphics[width=0.24\textwidth]{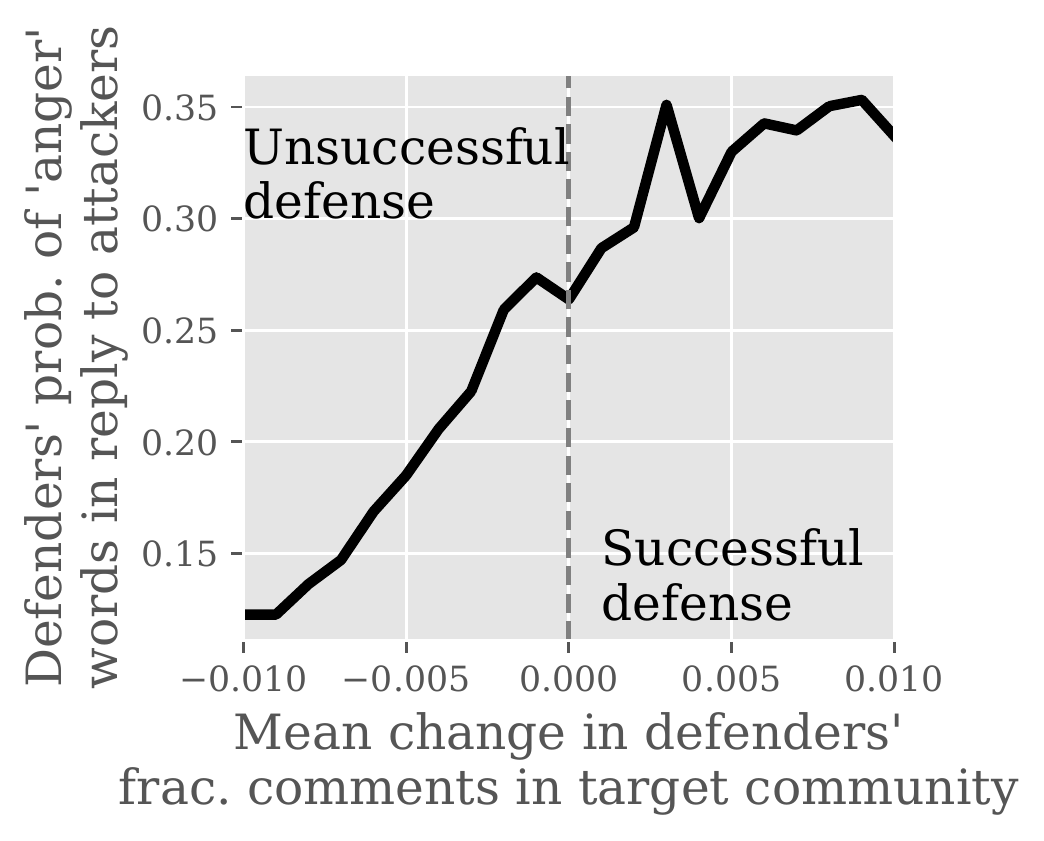}
        }
        \vspace{-10pt}
    \caption{(a) Two reply networks on two real target threads from our data showing key characteristic differences between successful and unsuccessful defense networks. In an unsuccessful defense, attackers ``gang-up'' on defenders, while defenders engage directly with attackers in successful defenses. In successful defenses: (b) defenders reply more to attackers, (c) defender's A-PageRank is higher, (d) attacker's D-PageRank is higher, and (e) defenders use more angry words in replies to attackers.}
    \label{fig:success}
    \vspace{-10pt}
\end{figure*}

\vspace{-1mm}
\subsection{Quantifying impact}

We measure the impact of mobilization after it is over as the change in attackers' and defenders' participation in the target community.
All changes are measured as differences between average activity levels of the users in the 30 days before mobilization starts and in the 30 days after the mobilization ends (measured as 3 days after the initial cross-link post was created).
As a baseline, we again compare with randomly matched comparison users (see Section \ref{sec:data} for details).

In Figure~\ref{fig:impact}, we measure the change (after minus before) in the fraction of comments that attackers and defenders make in the target community after negative mobilizations end (with analogous values for neutral mobilizations shown). 
A positive (negative, \textit{resp.}) change indicates an increased (decreased, \textit{resp.}) engagement in the target community.
We see that after negative mobilizations, attackers post more frequently in the target community while defenders post less frequently ($p<0.001$), indicating that negative mobilizations often lead to ``colonization'', where members of the attacking community become regular members of the target community. 
This colonization may lead to further adverse effects for the target community, since these incoming users are generally angrier than average and tend to violate community norms (\cf, Section~\ref{sec:interaction}).  

In contrast, neutral mobilizations lead to ``immigration'', where the mobilized users become more active in the target community, with no significant change for the `defenders' (Figure \ref{sec:impact}).
Moreover, as mobilized users in neutral mobilizations are more civil (\cf, Section~\ref{sec:interaction}), this effect could lead to an increase in positive cross-community  exchanges, which is an open area for future research.

\vspace{-1mm}
\subsection{What makes a defense successful?}
As negative mobilizations adversely affect the target community, are there any possible ways of preventing this from happening?
Is ignoring attackers the optimal way, as in the case of anti-social behavior such as trolling~\cite{binns2012don}, or is direct engagement more effective, as suggested in hostility reduction theories~\cite{allport1979nature,pettigrew1997generalized,pettigrew2000does}?

In order to understand strategies that may be useful for mitigating these adverse impacts of negative mobilizations, we compare cases of successful and unsuccessful defense. 
We define a defense to be `successful' when defenders become more active in their community after the negative mobilization ends, while an `unsuccessful' defense is when the defenders become less active.
\footnote{Alternate definitions of `success', \eg, those involving quality of future comments, can be explored in future work.}
We compare the 10\% most and 10\% least successful defenses and report differences in mean statistics between the two classes. 

Figure~\ref{fig:success}(a) illustrates the largest connected components of two real reply networks from our data---one for an unsuccessful defense and another for a successful defense, with several observations. 
The accompanying Figures~\ref{fig:success}(b--e) show characteristic differences between successful and unsuccessful defenses overall.
The x-axis in these figures varies the mean change (after minus before) in defender's proportion of comments in the target community after the negative mobilization ends---positive values indicate increased proportion (i.e., successful defense) and negative values indicate a decrease (i.e., unsuccessful defense), relative to the expected baseline change (calculated from the matched users defined in Section~\ref{sec:data}). The plots are smoothened as moving average (window size of $\pm$5) to remove minor variations.
Together, these figures show that a defense is successful when defenders tend to engage in a direct fierce dialogue with the attackers, thus preventing them from ``ganging-up'' on defenders, and breaking echo chambers that generally form during negative mobilization (see Section~\ref{sec:interaction}).
We explain the figures in detail in the next few paragraphs.

While there is no statistically significant increase in the number of defenders ($p=0.05$), we find that successful defenses are marked by a strong increase in fraction of defenders' replies towards attackers (Figure~\ref{fig:success}(b); correlation coefficient $=0.97$).
Further, we use the two novel PageRank variants, A-PageRank and D-PageRank, developed in Section~\ref{sec:interaction}.
Successful defenses have defenders with higher A-PageRank (0.036 \vs\ 0.031, $p<0.001$; Figure~\ref{fig:success}(c)) and attackers with higher D-PageRank (0.052 \vs\ 0.028; $p<0.001$; Figure~\ref{fig:success}(d)).
Thus, defenders facilitate more discussions with attackers in successful defense and there is significantly more cross-communication between them, preventing the formation of echo chambers.

But does this increased cross-communication entail a civil discussion or a fierce debate?
We answer this by calculating the probability of users using `anger' words (from LIWC dictionary~\cite{tausczik_psychological_2010}) in direct replies by attackers to defenders, and vice-versa. 
We find that in successful defenses, defenders use more anger words towards attackers than attackers do to defenders (0.017 \vs\ 0.015; $p<0.05$), while in unsuccessful defenses, both are equally angry towards each other ($p=0.37$).  
This shows that defenders become more aggressive towards attackers in cases of successful defense.

\vspace{2mm}
Overall, we find that negative mobilizations often lead to ``colonization'', where after the conflict is over, attackers become regular members of the target community, while defenders become less active there. 
However, we find that this negative outcome is prevented when defenders break echo-chambers and 
engage in direct heated conversations with attackers.

\vspace{-1mm}
\section{Prediction of Mobilizations}
\label{sec:predictions}

With the analysis in the previous sections, we have shown that negative mobilizations can have long-term adverse impact. 
So, can we predict whether a post will lead to a negative mobilization as soon as it is created?
An efficient model can then be used to create an early warning system for community moderators to take appropriate precautions.
In this section, we focus on this prediction task.

We create a novel ``socially-primed'' LSTM model along with a feature-based baseline for comparison.
Our deep learning model leverages both the text of the post as well as the user-to-community interaction network for predictions. 
We show that our model significantly outperforms the feature-based baseline for the task of predicting whether a cross-linking post will mobilize users. 

\begin{figure}
\centering
\includegraphics[width=0.3\textwidth]{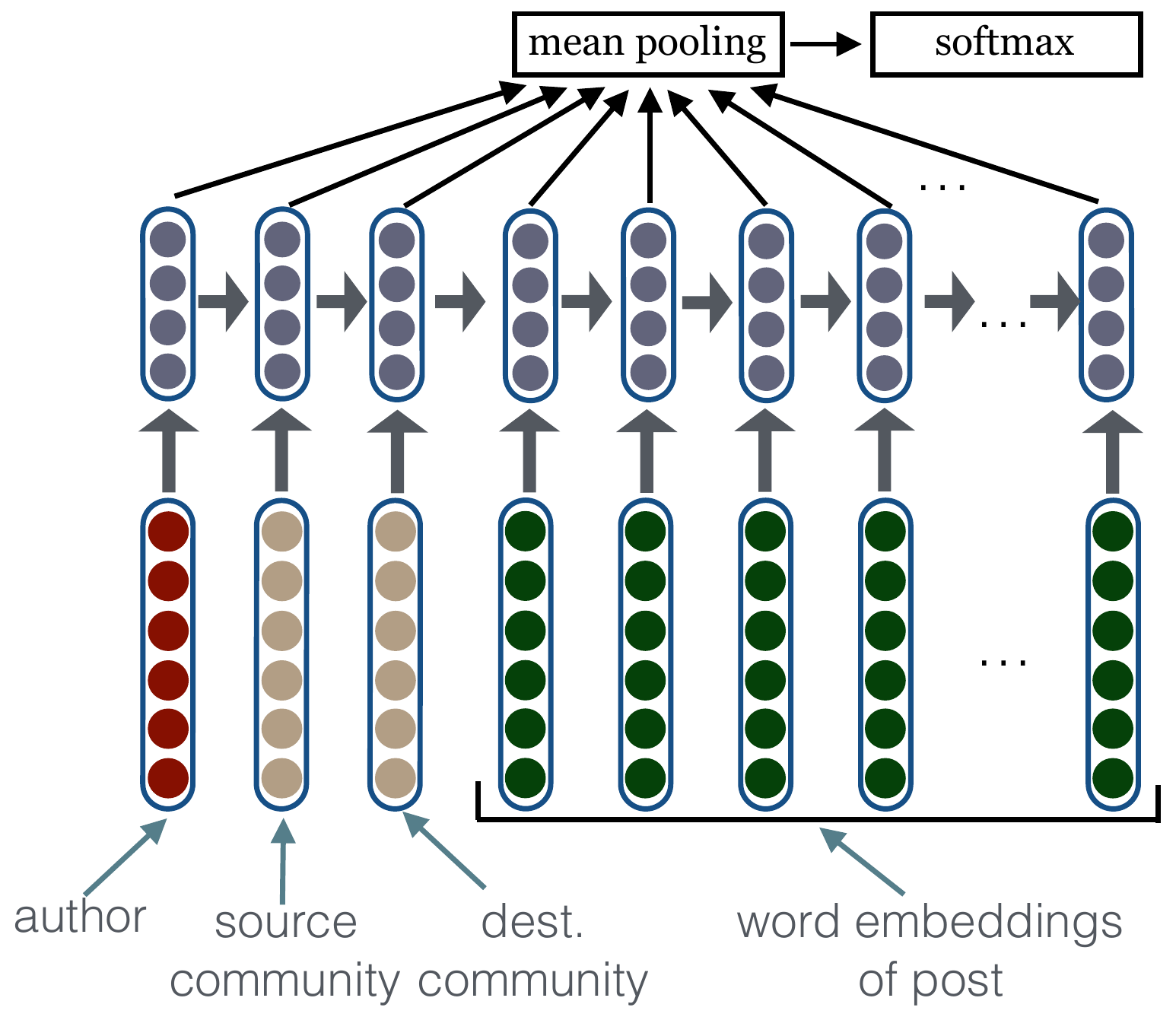}
\vspace{-10pt}
\caption{Socially-primed LSTM architecture.\label{fig:sp}}
\vspace{-10pt}
\end{figure}

\vspace{-2mm}
\subsection{Feature-based baseline model}
\label{sec:features}
As a baseline, we create a traditional feature based model, with a large set of 263 hand-crafted features, including\footnote{The full feature list is available in the project website~\cite{appendix}.}
\begin{itemize}[topsep=0pt, leftmargin=10pt]
\item
Post features: Lexical count statistics from the post (\eg, from LIWC \cite{tausczik_psychological_2010} and from VADER \cite{hutto2014vader}) and tf-idf similarity values between the post content of the source/destination communities. 
\item
User features: Activity levels of the post creator (\eg, fraction of posts in target) and averaged lexical features of the user's previous posts. 
\item
Community features: tf-idf similarities between source and target content.
\end{itemize}
These features are used both as a baseline in a Random Forest model~\cite{pedregosa_scikit_2011} and combined together with the deep model as an ensemble, as explained later.

\vspace{-2mm}
\subsection{Text, user, and community embeddings}
While it is common for deep learning models to use textual information, our data also contains rich social information about users and communities, which we can leverage in our predictions. 
Therefore, we convert text, user, and communities into 300-dimensional embeddings, which will be used as input to our deep learning model.
To generate text embeddings, we use standard techniques.\footnote{Off-the-shelf GloVe word embeddings trained on 840 billion words \cite{pennington2014glove}.}  
To generate user and community embeddings, we innovatively use the `which user posts in which community' network to convert users and communities into embedding vectors with the following intuition: two community embeddings will be close together if similar users post on them, and two user embeddings will be similar if they tend to post on the same communities.

More formally, we train user and community embeddings based on the bipartite multigraph between users and communities, where one user-community edge, $(u_i,c_j)$, exists in the graph for each time a user, $u_i$, posts on community, $c_j$. 
Adapting the approach used in a number of recent successful works  \cite{hamilton2017representation,mikolov_distributed_2013,perozzi2014deepwalk}, we learn user and community embeddings, $\mb{u}_i \in \mathbb{R}^d, \mb{c}_j \in \mathbb{R}^d$, from this graph using a negative-sampling optimization algorithm with the following loss:
{\small
\begin{equation}\label{eq:embeds}
\mathcal{L} = \sum_{(u_i,c_j) \in \mathcal{E}} -\log(\sigma(\mb{u}^\top_i\mb{c}_j)) - K\cdot\mathbb{E}_{c_n \sim P_n}\log(-\sigma(\mb{u}_i^\top\mb{c}_{n})),
\end{equation}
}
where $\E$ is the set of edges in the bipartite multigraph, $P_n$ is a uniform distribution over all communities (used to generate negative samples), and $K=5$ is the number of negative samples used. 

We trained the embeddings using 100 passes over the full 40 months of Reddit posts, with stochastic gradient descent code adapted from Levy et al. \cite{levy2014dependency}. 
Figure~\ref{fig:tsne} shows the resulting community embeddings plotted in a two-dimensional visualization using t-SNE \cite{maaten2008visualizing}, with communities sized according to the number of posts that are made in it, and colored as the proportion of its outgoing cross-links that are negative (red indicates high negativity, and blue indicates low negativity).
Communities were grouped manually into high-level categories, such as sports teams, technology, etc.
We observe that most community clusters are mostly blue, while controversial topics (\eg, cluster with r/conspiracy and r/theredpill) and advice communities generate  a high number of negative cross-links.

These text, user, and community embeddings will be used as input to our socially-primed LSTM model explained below.

\vspace{-2mm}
\subsection{Socially-primed LSTM model}
To combine social and textual data in a structured manner for predictions, we create a novel ``socially primed'' LSTM model. 
Using the word, user, and community embeddings as inputs, we train a recurrent LSTM model, as shown in Figure~\ref{fig:sp}.
LSTMs are a popular variant of recurrent neural networks (RNNs), \ie, neural networks that operate on sequential input sequences, $[\mb{x}_1, ..., \mb{x}_T]$ \cite{hochreiter1997long}.
The basic idea behind LSTMs is that they use a combination of dense neural networks to map the current input value, $\mb{x}_t \in \R^{d}$, along with a hidden-state vector, $\mb{h}_{t} \in \R^{h}$, to a new hidden state, $\mb{h}_{t+1} \in \R^{h}$. 
  
In our socially-primed LSTM model, the input sequences are the concatenation of (i) the user embedding of the post author, $\mb{u}_i$, (ii) the community embeddings of the source and target communities, $\mb{c}_s$ and $\mb{c}_t$, and (iii) the word embeddings of the post text, $[\mb{w}_0, ..., \mb{w}_L]$ (where $L$ denotes the length of the text).\footnote{We use a maximum length of $L=50$, discarding words that occur after this point.}
  Traditional LSTMs work with text data only, while `social-priming' signifies that user and community information are additionally used for training.
To predict mobilization, we take mean of the hidden LSTM states from each time step and then feed the resulting vector through a softmax (i.e., logistic) layer:
{\small
\begin{align}
[\mb{h}_1, ..., \mb{h}_{L+3}] &= \textrm{LSTM}([\mb{u}_i, \mb{c}_s, \mb{c}_t, \mb{w}_1, ..., \mb{w}_L])\\
y &= \frac{1}{1 + \exp\left(\boldsymbol{\theta}^\top (\frac{1}{L+3}\sum_{t=1}^{L + 3}\mb{h}_t\right)},
\end{align}
}
where 
$y$ denotes the probability of the post leading to mobilization, and $\mb{h}_t$ are the LSTM hidden states.\footnote{We trained the model using PyTorch~\cite{pytorch}, the Adam optimizer~\cite{kingma2014adam}, and a standard cross-entropy loss. 
The project website~\cite{appendix} contains data and code for replication.}

\vspace{-1mm} 
\subsection{Prediction results}
The prediction task we consider is: whether a post will mobilize source members, using the full set of cross-links 
from Section~\ref{sec:data}. 
We randomly use 80\% of the data for training, 10\% for validation, and the remaining 10\% as held-out test set.\footnote{We do not restrict to negative mobilizations, since the sentiment labels themselves were generated by a machine learning classifier. Predicting negative mobilization can be done by coupling this classifier with the sentiment classifier developed in Section~\ref{sec:data}.}

In this task, the Random Forest feature-based model (size 500) achieves an AUC of 0.67.
For reference, we include the traditional LSTM model that only uses text information (\ie, no user and community embeddings), which achieves an AUC of 0.66, almost equal to the random forest model.
Our socially-primed LSTM model significantly outperforms both these models, with an AUC of 0.72. 
We also create a Random Forest ensemble model that uses all the hand-engineered features, user embeddings, community embeddings, and the average hidden state of the LSTM, further improving performance to an AUC of 0.76.
This shows that both features and the LSTM model contribute towards the improved performance of the ensemble.

\vspace{2mm}
Overall, in this section, we created several machine learning prediction models and achieve high AUC of 0.76 for predicting if a post will mobilize users. 
These models can be used in practice to create early warning systems for community moderators to predict potential negative mobilizations as soon as the cross-link post is created, and help them to curb the adverse effects of these events.

\vspace{-1mm}
\section{Further Related Work}
\label{sec:related}

In addition to the computational social science and social psychology literature previously mentioned, our work also draws on and contributes to a rich line of research on online hate speech and anti-social behavior.
Research on negative interactions in online platforms has focused primarily on controversies~\cite{addawood2017telling,conover2011political,garimella2016quantifying,garimella2017reducing,lamba2015tempest,mejova2014controversy} and anti-social behavior, in the form of trolling~\cite{cheng2015antisocial,cheng2017anyone,kumar2014accurately}, sockpuppetry~\cite{kumar2017army}, harassment and cyberbullying~\cite{blackburn2014stfu,golbeck2017large,hinduja2014bullying,hosseinmardi2014towards,munger2017tweetment,ratkiewicz2011detecting,wulczyn2017ex},  vandalism~\cite{kumar2015vews}, hate speech~\cite{burnap2016us,djuric2015hate,matias2015reporting,nobata2016abusive,silva2016analyzing}, and others \cite{ferrara2017contagion,guntuku2017detecting,kumar2016disinformation,kumar2018rev2,kumar2018survey,lee2013crowdturfers,mitra2015credbank,saif2016role,wang2014cursing,yasseri2012dynamics}.
While these studies cover a broad spectrum of anti-social behavior, they focus on user-level analyses. In contrast, our study focuses on interactions at the community-level.
Recent research has also shown that organized negative behavior can spread across platforms~\cite{hine2017kek,zannettou2017web}, and that banning communities that foster hate-speech was effective in decreasing hate speech on Reddit~\cite{chandrasekharan2017you}.
Unlike these studies, which give important insights into the dangers of online negative behavior and effectiveness of correction measures, our research focuses on (negative) interactions between community-community dyads and their impact.

\vspace{-3mm}
\section{Discussion and Conclusions}
\label{sec:conclusion}

Our findings provide a novel, large-scale view of intercommunity interaction and conflict on the web.
By leveraging cross-links between posts made in communities and carefully controlling for baseline rates of user activity, we were able to identify explicit cases of intercommunity mobilization.

Our analysis shows that negative mobilizations have important long-term adverse effects, leading to processes of ``colonization'', where ill-behaved users come to dominate the target community. 
However, we also uncovered mitigating factors that correlate with decreased rates of these adverse outcomes---most prominently, that increased engagement in discussions between members of the interacting communities leads to improved outcomes. 
Further, we designed a novel socially-primed LSTM model, which combines user, community, and text embeddings, to predict whether mobilizations will occur with an AUC of 0.76.

Nonetheless, our analysis has limitations that point to interesting directions for future work.
First, we study interactions between pairs of communities, but understanding interactions between more than two communities could reveal larger-scale dynamics of intercommunity interactions.
Our analysis is also based on a single platform where users are pseudonymous, while the nature of community interactions and conflicts may be different where people use their `real' identities (\eg, Facebook or Twitter).
Moreover, while we study the impact of mobilization on the involved users, an analysis of conversations in the rest of the community could further our understanding of the impact of intercommunity interactions.
Additionally, the relation between intercommunity interactions and anti-social behavior, such as trolling~\cite{cheng2015antisocial} and sockpuppetry~\cite{kumar2017army}, is yet unexplored.
Lastly, our analysis of the sentiment of intercommunity mobilizations is limited to instances of {\em explicit sentiment}, where the negative (or neutral) sentiment is clear to a relatively uninformed crowdworker. 
However, community restrictions may prevent creation of explicit derogatory posts and lead to instances of {\em implicit negative sentiment}.
Detecting implicit sentiment and using it to study intercommunity interaction is an important open direction for future work.

By identifying, analyzing, and predicting intercommunity mobilizations, our work outlines a data-driven methodology for quantifying the impact of these interactions and predicting when they will occur. 
The management of highly negative and disruptive communities in particular is becoming increasingly important for multi-community platforms~\cite{nytimes}, and our analysis helps to pave the way towards the development of policies, strategies, and tools for promoting positive interactions on these platforms.

\vspace{-2mm}
{\small
\begin{acks}
Parts of this research are supported by NIH BD2K, DARPA NGS2, ARO MURI, IARPA HFC, ONR CEROSS, Stanford Data Science Initiative, Chan Zuckerberg Biohub, IIS-1514268, SAP Stanford Graduate Fellowship and NSERC PGS Doctoral grant. The authors thank Dr. Ajay Divakaran and Dr. Karan Sikka for helpful discussions.
\end{acks}
}

\bibliographystyle{abbrv}
\balance
\bibliography{references}

\end{document}